\documentclass[conference]{IEEEtran}
\IEEEoverridecommandlockouts
\usepackage{cite}
\usepackage{amsmath,amssymb,amsfonts}
\usepackage{algorithmic}
\usepackage{graphicx}
\usepackage{textcomp}
\usepackage{xcolor}
\usepackage{booktabs}
\usepackage{tabularx}
\usepackage{blindtext}
\usepackage{hyperref}
\def\BibTeX{{\rm B\kern-.05em{\sc i\kern-.025em b}\kern-.08em
    T\kern-.1667em\lower.7ex\hbox{E}\kern-.125emX}}

\makeatletter
\def\ps@headings{
\let\@oddhead\@empty
\let\@evenhead\@empty
\def\@oddfoot{\@IEEEheaderstyle\hfil\thepage}%
\def\@evenfoot{\@IEEEheaderstyle\thepage\hfil\hbox{}}
}
\def\ps@IEEEtitlepagestyle{
\let\@oddhead\@empty
\let\@evenhead\@empty
\def\@oddfoot{\footnotesize ©2024 IEEE\hfill}%
\let\@evenfoot\@empty
}
\makeatother

\begin{document}

\title{Cybersecurity in Industry 5.0: Open Challenges and Future Directions\\}

\author{\IEEEauthorblockN{Bruno Santos}
\IEEEauthorblockA{\textit{CIIC, ESTG} \\
\textit{Polytechnic University of Leiria}\\
Leiria, Portugal \\
bruno.b.santos@ipleiria.pt}
\and
\IEEEauthorblockN{Rogério Luís C. Costa}
\IEEEauthorblockA{\textit{CIIC, ESTG} \\
\textit{Polytechnic University of Leiria}\\
Leiria, Portugal \\
rogerio.l.costa@ipleiria.pt}
\and
\IEEEauthorblockN{Leonel Santos}
\IEEEauthorblockA{\textit{CIIC, ESTG} \\
\textit{Polytechnic University of Leiria}\\
Leiria, Portugal \\
leonel.santos@ipleiria.pt}
}

\maketitle

\begin{abstract}

Unlocking the potential of Industry 5.0 hinges on robust cybersecurity measures. This new Industrial Revolution prioritises human-centric values while addressing pressing societal issues such as resource conservation, climate change, and social stability. Recognising the heightened risk of cyberattacks due to the new enabling technologies in Industry 5.0, this paper analyses potential threats and corresponding countermeasures. Furthermore, it evaluates the existing industrial implementation frameworks, which reveals their inadequacy in ensuring a secure transition from Industry 4.0 to Industry 5.0. Consequently, the paper underscores the necessity of developing a new framework centred on cybersecurity to facilitate organisations' secure adoption of Industry 5.0 principles. The creation of such a framework is emphasised as a necessity for organisations.

\end{abstract}

\begin{IEEEkeywords}
Cybersecurity, Industry 5.0, Industrial Frameworks, Cyberattacks, Countermeasures
\end{IEEEkeywords}

\section{Introduction}

Recent Industrial Revolutions have led to a significant increase in the digitalisation of industrial processes \cite{Leng2022}. With the widespread adoption of industrial digital systems, it is of great importance to implement robust cybersecurity measures to safeguard information, assets, and individuals. Failure to do so can have a significant impact on human beings, both physically and mentally. For instance, cyberattacks that alter the typical behaviour of collaborative robots can result in physical damage to the human and the robot, as well as the products \cite{Hollerer2021,Jia2022}. Furthermore, privacy data breaches can impact negatively the mental health of the people involved \cite{mental_health1}. A recent report on data breaches analysed 30,458 security incidents in 2023, of which 10,626 were confirmed data breaches \cite{Verizon2024}. This number of data breaches was a record high. In the manufacturing industry, 2,305 incidents were analysed, of which \(\approx\) 37 \% had confirmed data breaches. Furthermore, the compromised data in the data breaches was mainly personal data. These alarming numbers show that security incidents affect not only the organisations but also the employees and customers. In Industry 5.0, organisations must prioritise strong cybersecurity measures to protect all the people and assets involved.

This paper examines potential cyberattacks and corresponding countermeasures of Industry 5.0 technologies, demonstrating the new threats they bring to organisations. Also, the paper examines current industrial implementation frameworks to address whether there is a necessity for the creation of a new framework for the secure implementation of Industry 5.0. The analysed frameworks were: Industrial Internet Reference Architecture (IIRA), Reference Architecture Model for Industry 4.0 (RAMI 4.0), Guide to Operational Technology (OT) Security by the National Institute of Standards and Technology (NIST), cybersecurity guides by the Instituto Nacional de Ciberseguridad (INCIBE), and two academic frameworks. The frameworks were found to be inadequately prepared for the transition to Industry 5.0 because they do not address concerns about cybersecurity, the green transition, human well-being and hyper customisation. These concerns are the pillars of Industry 5.0. As a result, the paper emphasises the importance of developing a new framework focused on cybersecurity to enable organisations to adopt Industry 5.0 principles securely.

The contributions of this paper are summarised as follows:
\begin{itemize}
    \item Identification of Industry 5.0 enabling technologies and a review of the literature to gather information about cyberattacks on enabling technologies and their respective countermeasures. This review gathers the current knowledge on the attack surfaces of these technologies.
    \item Analysis of the current most known and complete frameworks for Industry 4.0 and Industry 5.0. The main goal of the analysis is to test the use of these frameworks in the transition from Industry 4.0 to Industry 5.0. The analysis will focus on the cybersecurity aspects of these frameworks, as the transition to Industry 5.0 depends heavily on it.
    \item Identification of open challenges and future work directions. 
    
\end{itemize}

The paper is organised as follows. Section \ref{sec:background} provides detailed background information and reviews related work in the literature. Section \ref{sec:challenges_frameworks} reviews potential cyberattacks on the new enabling technologies of Industry 5.0 and their respective countermeasures. Furthermore, it analyses current industrial frameworks for the implementation of Industry 5.0. Finishing with the open challenges and future directions inherent in this theme. Finally, the paper concludes with Section \ref{sec:conclusion}.

\section{Background and Related Work}
\label{sec:background}
This section begins by reviewing essential background information related to the different industrial revolutions and the major differences between the latest industrial revolutions, Industry 4.0 and Industry 5.0. Furthermore, it points out the key enabling technologies of Industry 5.0, followed by a review of related work.

Many historians, economists, and scholars define industrial revolutions as periods of technological change with a high impact on society \cite{Klingenberg2022}. As of the writing of this paper, there have been five known industrial revolutions. In the 1800s, the First Industrial Revolution, also known as Industry 1.0, developed mechanical production infrastructures for water and steam-powered machines. Industry 2.0, the Second Industrial Revolution, emerged in 1870 with the introduction of electric power and assembly line production. Industry 3.0 came into being in 1969, with the introduction of electronics, partial automation, and Information Technologies (IT) \cite{Maddikunta2022}. Industry 4.0 evolved in 2011 with the concept of smart manufacturing by merging IT and Operational Technology (OT) in a cyber-physical system (CPS) to achieve mass automation and production \cite{Maddikunta2022,Leng2022}. Because Industry 5.0 is still in its initial stages different definitions are being provided by industry practitioners and researchers. This paper combines the definitions provided by \cite{EuropeanComission2022} and \cite{Maddikunta2022}. The definition is as follows: Industry 5.0 aims to achieve societal goals beyond efficiency and productivity, transforming industries into resilient providers of prosperity. For that objective, industries must respect the planet and lead the green transitions. It also places the well-being of the industry worker at the centre of the production process and instead of replacing humans with machines, a collaboration between both takes place. This collaboration is designed to use the creativity of human experts who work together with efficient, intelligent and accurate machines. Furthermore, the new technologies allow for hyper customisation, providing customers with more specific products, services and content. Industry 5.0 complements Industry 4.0 by investing in the transition to a more sustainable, human-centric, hyper customisable and resilient industry. \autoref{tab:comparison} presents the differences between Industry 4.0 and Industry 5.0 in terms of key components and their objectives.

\renewcommand{\arraystretch}{1.5}
\begin{table}[!htpb]
    \centering
    \caption{Comparison Between Industry 4.0 and Industry 5.0 Key Components and Objectives.}
    \label{tab:comparison}
    \scriptsize
    \begin{tabular}{p{0.15\linewidth}p{0.3\linewidth}p{0.43\linewidth}}
    \toprule
     & \begin{tabular}[c]{@{}c@{}}\textbf{Key Components}\end{tabular} & \begin{tabular}[c]{@{}c@{}}\textbf{Objective}\end{tabular} \\
    \midrule
        \textbf{Industry 4.0} & Automation, cyber-physical systems and smart manufacturing.& Increase productivity and decrease costs in production. \\
        \textbf{Industry 5.0} & Human-machine collaboration, green transition, human creativity and well-being, and hyper customisation. & Bring humans back to the manufacturing process. Increase well-being and job satisfaction. Increase social stability. Transition to a more sustainable and resilient industry. Allow customers to customise specific products, services and content. \\
    \bottomrule
    \end{tabular}
\end{table}

\subsection{Enabling Technologies for Industry 5.0}
\label{sec:technologies}
In 2020, the Directorate-General for Research and Innovation of the European Commission wrote a report listing the enabling technologies for Industry 5.0 \cite{DGRIC2020}. This report was written based on a workshop with some of Europe’s technology leaders. Other papers like \cite{Maddikunta2022} and \cite{Leng2022} also reference some of the same enabling technologies. This section will specifically concentrate on the European Commission report as it provides more detailed information.

\begin{figure}[!htpb]
    \centering
    \includegraphics[width=.85\linewidth]{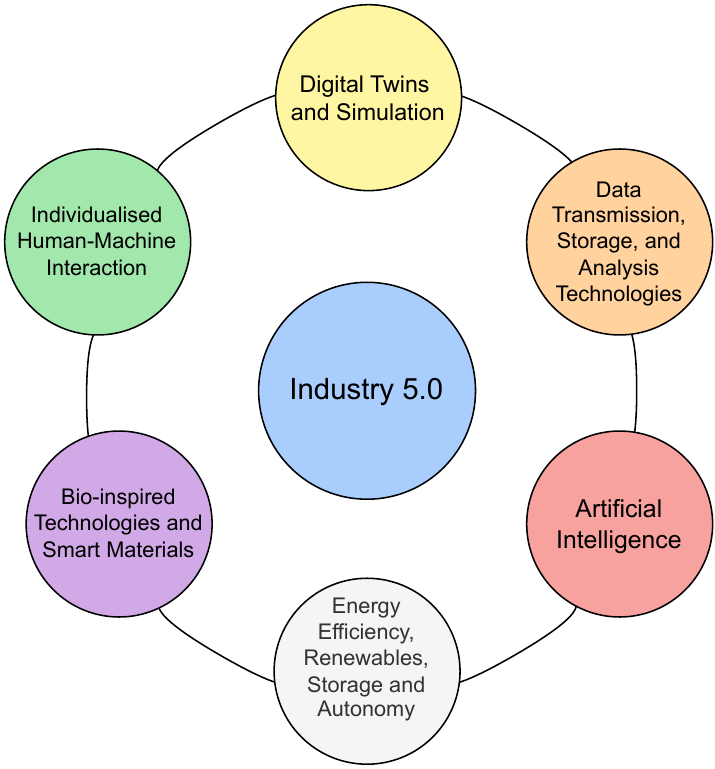}
    \caption{Categories of Industry 5.0 Enabling Technologies.}
    \label{fig:enabling_technologies}
\end{figure}

The report lists a total of 41 enabling technologies and properties. Each enabling technology is classified into one of six different categories, as can be seen in \autoref{fig:enabling_technologies}. Due to the numerous enabling technologies, a random selection was made to avoid unnecessary lengthening of this subsection. For the full list of enabling technologies, please refer to the report mentioned.

In the category of individualised human-machine interaction, there are technologies like collaborative robots, also known as cobots, exoskeletons and mental and physical strain sensors. These technologies help increase the well-being of workers and job satisfaction and bring humans back to the manufacturing process.

In the category of bio-inspired technologies and smart materials, technologies like living and recyclable materials are essential for Industry 5.0. These technologies will help with the green transition, building a more sustainable and resilient industry.

In the category of digital twins and simulation, there are technologies like simulations and digital twins of entire systems, and simulations of environmental and social impact. These technologies allow customers to personalise specific products, services and content, and help with the transition to a more sustainable and resilient industry.

In the category of data transmission, storage, and analysis, technologies that help with cybersecurity and safe cloud infrastructure, big data analytics, traceability and edge computing are essential. These technologies help towards the hyper customisation of products, services and content, and help build a more resilient industry.

Technologies in the artificial intelligence (AI) category, like brain-machine interfaces, human-centred machine learning and explainable AI are also crucial. These technologies increase well-being and job satisfaction, allow hyper customisation and bring humans back to the manufacturing process.

In the category of technologies for energy efficiency, renewables, storage and autonomy, the integration of renewable energy sources and energy-autonomous sensors is also crucial for a green transition in Industry 5.0. These technologies will help the transition to a more sustainable and resilient industry.

\subsection{Related Works}

Numerous researchers have studied the enabling technologies of Industry 5.0, their roles in the industrial revolution, and the security implications of their implementation. This subsection intends to showcase various works related to the topics at hand.

In 2021, Farsi et al. \cite{Farsi2021} identified possible enabling technologies of Industry 5.0. The authors also created a framework that works as a roadmap for the implementation of the enabling technologies in the short, medium and long term. The roadmaps also included cultural and organisational goals. The framework underwent validation via a comprehensive literature review and surveying a diverse group of participants from various industrial sectors.

In 2022, Maddikunta et al. \cite{Maddikunta2022} discussed the key enabling technologies of Industry 5.0 and their possible application and potential, and presented security and privacy challenges for the future. The future challenges identified by the authors were: security, privacy, human-robot co-working in factories, scalability, lack of skilled workforce and compliance with regulations. 

In the same year, Leng et al. \cite{Leng2022} presented key enabling technologies, discussed security and privacy challenges, and built a tri-dimension system architecture for the implementation of Industry 5.0. The authors express their privacy and security concerns by saying that the human-centric Industry 5.0 will generate a lot more data related to humans, posing challenges to the cybersecurity of this industrial revolution. The architecture proposed for the implementation of Industry 5.0 by the authors is divided into enablers, implementation path and applicability.

In a recent work in 2024, Hassan et al. \cite{Hassan2024} discussed the risks and mitigations of the adoption of Industry 5.0. The authors started by identifying the enabling technologies for Industry 5.0, followed by the identification of risks and countermeasures. The identification of risks and mitigations was done based on a review of other literature. Furthermore, the authors also categorise the risks into cybersecurity risks, workforce and training risks, operational and implementation risks, and other risks.

Our literature review covers key enabling technologies, security and privacy challenges, cyberattacks and countermeasures, and frameworks for implementing a secure Industry 5.0.

\section{Cybersecurity Challenges in Industrial Frameworks}
\label{sec:challenges_frameworks}

This section examines potential cyber threats targeting the emerging technologies of Industry 5.0, along with corresponding mitigation strategies. Additionally, it analyses existing industrial frameworks on the adoption of Industry 5.0, concluding with an exploration of the open challenges and future directions. These topics will be divided into their respective subsection.

\subsection{Cyberattacks and Countermeasures in Industry 5.0}
\label{sec:attacks}

For every enabling technology previously identified, an analysis of potential cyberattacks and corresponding countermeasures will be conducted in this section. The purpose of this analysis is to assess the security risks that are associated with each Industry 5.0 enabling technology and to provide effective measures to mitigate them. The cyberattacks and countermeasures presented will be based on a review of the current literature.

\autoref{tab:cyberattacks_pt1} displays the enabling technology, associated cyberattacks and respective countermeasures. Not all enabling technologies mentioned in the previous section are in the table because some do not have any cyber applications or known cyberattacks, such as recyclable and living materials. It is also worth mentioning that to focus only on the new enabling technologies of Industry 5.0, some technologies already in Industry 4.0, like simulations, digital twins, big data analytics and blockchain, will not be in this table.

As can be seen in \autoref{tab:cyberattacks_pt1}, multiple possible attacks and countermeasures in these technologies have already been studied. The most common attack is the Denial of Service (DoS), which occurs in almost all technologies. The Man-in-the-Middle (MitM) attack has also been identified multiple times in various technologies. In contrast to the potential cyberattacks, the countermeasures appear to be unique for each Industry 5.0 enabling technology. Accordingly, the new enabling technologies of Industry 5.0 increase the cybercriminals' attack surface. This can hold back organisations from transitioning to Industry 5.0. Secure implementation frameworks could help these organisations take another step further in the transition. These frameworks must address cybersecurity issues, which is a key element of this human-centred transition.

\renewcommand{\arraystretch}{1.5}
\begin{table*}[!htpb]
    \centering
    \caption{Review of Literature on Cyberattacks and Countermeasures of Industry 5.0 enabling technologies.}
    \label{tab:cyberattacks_pt1}
    \scriptsize
    \begin{tabular}{p{0.2\linewidth}p{0.3\linewidth}p{0.4\linewidth}}
    \toprule
    \begin{tabular}[c]{@{}c@{}}\textbf{Enabling Technologies}\end{tabular} & \begin{tabular}[c]{@{}c@{}}\textbf{Possible Cyberattacks}\end{tabular} & \begin{tabular}[c]{@{}c@{}}\textbf{Countermeasures}\end{tabular} \\
    \midrule
        Cobots & Physical tampering of data cables. Locally connected USB devices. DoS because of a shutdown button on the web application. Brute force of valid user names. Privilege escalation.  Exploiting Outdated Software (EOS). Cross-site scripting (XSS). MitM attack \cite{Hollerer2021}. & Physical access control. Apply the zero trust model. Removing the shutdown button from the web application. Usage of responses which do not indicate the existence of user accounts. Implementation of different access types. Updating the outdated software. Applying the missing HTTP headers \cite{Hollerer2021}.  \\
        
        Cobots (Cont.) & Modifying the controller parameters. Changing the calibration parameters. Modification of production logic. Change the status of the robot \cite{Raicu2022a}. Deliberate adversarial machine learning attacks \cite{Jia2022}. & Update firmware/software according to CVE. Update signatures on Intrusion Detection System (IDS). Policy intervention. Increase security awareness \cite{Raicu2022a}. Hand detection filter to prevent the cobot from causing physical damage to the hands. Use of a more robust machine learning model \cite{Jia2022}. \\
        
        Exoskeletons or Cobots that use Robot Operating System (ROS) & Unauthorised access of data via the subscription of topics in a ROS node. Publishing large amounts of data to a subscribed ROS node, creating a DoS attack \cite{Dieber2017,Breiling2017}. & Changing the network port of the ROS master node. Using TLS to secure the communications \cite{Dieber2017,Breiling2017}. \\
        
        Mental and physical strain sensors & MitM attacks. Overflow-based malicious code injection. Firmware attack \cite{Liu2016}. & Encrypted data payloads of packets. Revising the programming errors. Encrypt and decrypt the firmware on the wearable device with public key or symmetric key. Avoid wearing a smartwatch when typing confidential information \cite{Liu2016}. \\
        
        Mental and physical strain sensors (Cont.) & Replay attack. Parallel session attack. DoS attack. Stolen verifier. Server spoofing. Fake server. The leak of the verifier. Impersonation \cite{Soni2023}. & Put a timestamp on every message/data and introduce a threshold time within which data must be received. Transmit credentials with robust cryptographic techniques. Strong authentication protocols and mutual authentication. Server and firewall must be updated with the latest versions. Data stored in the database must be stored using a one-way hash function or other cryptographic method \cite{Soni2023}. \\
        
        Brain-machine interfaces & Side-channel attacks to reveal a user’s private information \cite{Martinovic2012}. Narrow period pulse attacks \cite{Meng2023}. Backdoor attacks \cite{Jiang2023}. & Not to expose the raw data from EEG devices to third-party applications. Adding noise to the EEG raw data before making it available to the applications that must use it \cite{Martinovic2012}. Fine-tuning. Stochastic activation pruning \cite{Meng2023}. Fine-tuning. Input preprocessing \cite{Jiang2023}. \\
        
        Explainable AI & Adversarial attacks on model explanations \cite{Baniecki2024}. Model extraction attack to extract the decision boundaries of white-box models such as decision trees and logistic regression \cite{Oksuz2023}. & Regularising a neural network. Aggregating multiple explanations created with various algorithms \cite{Baniecki2024}. Deep learning with differential privacy. Static distortion. Rounding the confidence values obtained during predictions \cite{Oksuz2023}. \\
        
        Energy-autonomous sensors (energy harvesting) & Eavesdropping. DoS attacks. Side channel attacks. Device Tampering. Replay attacks. Spoofing attacks. MitM attacks. Malware injection \cite{Tedeschi2020}. & Provide a framework that as more energy becomes available, authentication and confidentiality are strengthened. Use precomputation techniques that anticipate the execution of the most energy-demanding tasks when the device is fully powered \cite{Tedeschi2020}. \\
        
        Energy-autonomous sensors (energy harvesting) (Cont.) & Flooding attacks. Jamming attacks. DoS attacks at network level and stealthy collision attacks. Energy DoS attack. Power and timing side-channel attacks \cite{Tedeschi2020}. & Measuring throughput degradation under flooding attacks. Channel hopping, time splitting energy harvesting, fake transmissions. Adaptive acknowledgement approach. Power positive networking. Quantisation controllers \cite{Tedeschi2020}. \\
        
        Renewable energy sources & Meter fraud attacks \cite{Tang2023}. Adversarial learning attack against deep learning-based renewable energy forecasts \cite{Ruan2024}. Ramp attack \cite{Sarangan2018}. & Convolutional neural network-based detector \cite{Tang2023}. Use an anti-ramp attack algorithm \cite{Sarangan2018}. \\
        
        \\
    \bottomrule
    \end{tabular}
\end{table*}

\subsection{Existing Frameworks and Industry 5.0}
\label{sec:frameworks}

In this subsection, an analysis of Industry 4.0 implementation frameworks will be conducted to determine whether they can be applied to Industry 5.0 or if new frameworks are necessary to facilitate its implementation. Academic frameworks for Industry 5.0 will also be analysed briefly. The analysis will concentrate on the cybersecurity aspects of each framework but will not be limited to them. High-level Industry 4.0 frameworks will be analysed, namely  IIRA and RAMI4.0. Furthermore, two academic high-level frameworks for Industry 5.0 will be analysed briefly. Also, low-level architectures provided by NIST and INCIBE will be analysed. These architectures were chosen because they are the most studied and complete.

The IIRA provides a model for businesses to develop future products and business strategies by merging OT and Information Technology (IT) \cite{IIC2022}. The model focuses on the Industrial Internet of Things (IIoT) and is organised into four different Viewpoints: Business Viewpoint, Usage Viewpoint, Implementation Viewpoint and Functional Viewpoint. These Viewpoints are created to identify and classify the common preoccupations of an IIoT architecture. The Business Viewpoint identifies the participants involved in the system along with their business views, values, and objectives. The Usage Viewpoint focuses on the system's expected business outcomes. The Implementation Viewpoint looks at the technologies that are required to implement the functional components of the system and their communication schemes. Finally, the Functional Viewpoint examines the functional components of the system and how they interact with each other and with the external environment \cite{IIC2022}.

The RAMI 4.0 gives companies a framework for developing future products and business models, with the major goal of improving the manufacturing process through digitalisation \cite{PlattformIndustrie2018}. RAMI 4.0 consists of a three-dimensional coordinate system that describes all crucial aspects of Industry 4.0. The three axis are the Layers Axis, the Life Cycle \& Value Stream axis and the Hierarchy Levels axis. The Life Cycle \& Value Stream axis represents the life cycle of facilities and products, from the first idea to decommissioning. The Layers axis is divided into six layers each representing the decomposition of an asset into its properties. The Hierarchy Levels axis represents the flexible communication model, in which systems and machines can communicate across hierarchy levels \cite{PlattformIndustrie2018, Li2020}.

Leng et al. built a three-dimensional architecture for the implementation of Industry 5.0 \cite{Leng2022}. This architecture was made to stimulate discussion on the different components of Industry 5.0. Furthermore, the architecture is composed of the technical dimension, reality dimension, and application dimension. The technical dimension represents the enabling technologies. The reality dimension represents the implementation path. Finally, the application dimension represents the different application sectors of Industry 5.0 \cite{Leng2022}.

Aheleroff et al. also built a three-dimensional architecture, this time called RAMI 5.0 \cite{Aheleroff2022}. The architecture uses RAMI 4.0 as a base, which means that the three axis have the same names. The Life Cycle \& Value Stream axis represents the life cycle of hyper customisable products. The Layers axis is also the decomposition of an asset into its properties, but now addresses the sustainability, resilience, and cohesion of each asset. This axis is divided into physical and digital layers. The Hierarchy Levels axis is not detailed in the paper but it seems to represent the different applications of Industry 5.0 \cite{Aheleroff2022}.

The NIST Special Publication (SP) 800-82r3, provides a guide to OT Security \cite{Stouffer2023}. This guide provides secure architectures for different industrial control systems (ICS). These architectures are way more detailed and specific than IIRA or RAMI 4.0. Furthermore, the guide also provides cybersecurity best practices for many Industry 4.0 enabling technologies, such as IIoT and the Cloud \cite{Stouffer2023}.

INCIBE, a Spanish cybersecurity organisation, also provides various guidance in designing and configuring security architectures for ICS \cite{INCIBE}. This organisation separates their guides into multiple publications. These publications are detailed and cover some important security topics, such as intrusion detection/prevention systems and security information and event management systems \cite{INCIBE_SIEM}. Furthermore, they also talk about defence endpoints \cite{INCIBEDenfenceEndpoints} and how to do asset inventory management in an ICS architecture \cite{INCIBE_asset}.

Both IIRA and RAMI 4.0 are high-level reference architectures, which lack detailed information and are more general on how to implement Industry 4.0. The IIRA model includes cybersecurity concerns in almost every Viewpoint. However, the guidance on how to implement such cybersecurity measures is not detailed or explained. For example, the model says that security functions, such as encryption and authentication, are needed in every functional component but does not elaborate further. The model addresses users in the Usage Viewpoint, which is ideal for a human-centred Industry 5.0 implementation. However, the IIRA focus too much on IIoT systems, and Industry 5.0 is much more than that. Despite RAMI 4.0's much broader focus, it lacks even more detailed information on how to implement cybersecurity measures. Furthermore, this model does not address the human side of the manufacturing process. Neither of these models addresses the green transition nor the hyper customisation process. The two academic high-level frameworks for Industry 5.0 share a common deficiency: a failure to address cybersecurity. Despite not addressing cybersecurity, both these frameworks address the other Industry 5.0 major requirements. 

\begin{figure}[!htpb]
    \centering
    \includegraphics[width=.85\linewidth]{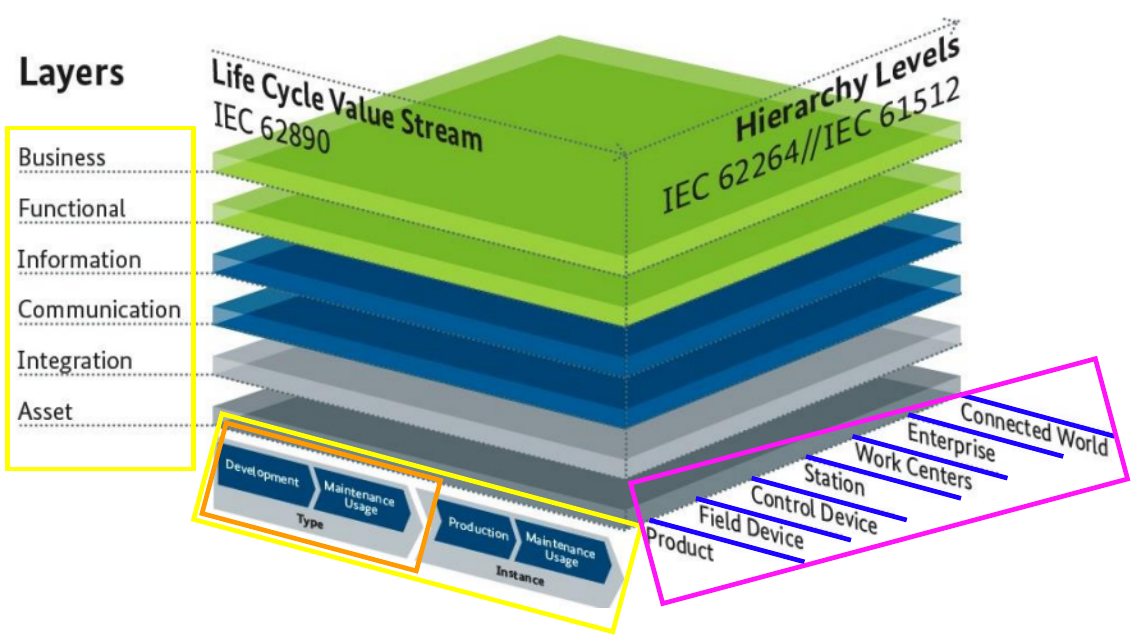}
    \caption{Highlight of RAMI 4.0's aspects that need modification.}
    \label{fig:rami_flaws}
\end{figure}

To illustrate the modifications that are required in these high-level architectures, an example will be presented using RAMI 4.0. \autoref{fig:rami_flaws} highlights the different aspects, of RAMI 4.0, that need to be modified to accommodate Industry 5.0. The yellow rectangles represent the aspects that need to be modified to accommodate the green transition. The Layers axis must represent properties such as sustainability and resilience of the products. Furthermore, the Life Cycle \& Value Stream axis must demonstrate the recyclability of the products during development and production. On the same axis, the orange smaller rectangle indicates the necessity for hyper customisation in the development of the products. On the Hierarchy Levels axis, cybersecurity, represented by the blue lines, should be present at every step of the hierarchy. Furthermore, the human being, represented by the pink rectangle, should be present in the hierarchy as well.

The NIST SP 800-82r3 and INCIBE's guide provide low-level architectures, which are more detailed and specific. The SP 800-82r3 guides the implementation of cybersecurity measures in various types of ICS architectures. However, the guide does not address or lacks detail in most Industry 4.0 and Industry 5.0 enabling technologies. Furthermore, the guide does not address the human side of the manufacturing process, which is essential for an Industry 5.0 implementation. Other key components like the green transition and the hyper customisation process are also not addressed. INCIBE's guides are scattered into multiple publications. These publications address the implementation of multiple cybersecurity measures. However, these guides do not seem as detailed as NIST's. INCIBE'S guides suffer from the same problems as NIST's guide.

In summary, the presented frameworks do not seem capable of helping organisations in the full process of transitioning to Industry 5.0. Accordingly, the creation of a new framework is needed to help this process.

\subsection{Open Challenges}
\label{sec:challenges}

Based on the analysis done in this paper, the following open challenges and areas for future research were identified:

\vspace{.2cm}\noindent\textit{Creation of High-level Secure Framework:} The high-level architecture should be abstract and more generalised, by focusing on deconstructing the complex topic of Industry 5.0. This architecture should explain the basic components and problems of Industry 5.0, making it more digestible to understand. By focusing on the basics of Industry 5.0, this architecture would make it more comprehensible and easier to digest for stakeholders.

\vspace{.2cm}\noindent\textit{Creation of Low-level Secure Framework:} At the lower level, various detailed and specific architectures should guide organisations. These architectures would represent the secure implementation of various enabling technologies in different industrial sectors and configurations. With multiple low-level architectures, a broader range of organisations can be reached while still being detailed and specific. This type of framework would help the organisation's technicians make the transition to Industry 5.0. 

\section{Conclusion}
\label{sec:conclusion}

Industries will have a crucial role in providing solutions for societal challenges, including resource conservation, climate change, and social stability. Industry 5.0 provides a vision beyond the improvement of efficiency and productivity seen in Industry 4.0. Industry 5.0 is human-centred, making cybersecurity crucial in the transition. Inappropriate cybersecurity measures can result in hazardous situations for workers or clients, or in the occurrence of significant privacy breaches. 

In this paper, an analysis of potential cyberattacks and countermeasures in the Industry 5.0 enabling technologies was made. With the increase in attack surface from all the new technologies, the implementation of Industry 5.0 needs to be secure. Accordingly, a review of current industrial frameworks was made, to test their capabilities for the safe implementation of Industry 5.0. The frameworks that were subjected to review were found to be unsuitable for the transition. Consequently, the paper emphasises the necessity of the creation of a new framework to assist organisations in securely transitioning to Industry 5.0, with cybersecurity as its foundation.

It is recommended for future work, to develop a framework capable of helping the secure transition to Industry 5.0. Furthermore, the framework should prioritise cybersecurity, human-machine collaboration, sustainable practices, human creativity, well-being, and personalised products and services. Also, testing must be done to ensure that the framework is validated.

\section*{Acknowledgment}
This work was mainly supported by the Sustainable Stone by Portugal agenda funded by the European Union/Next GenerationEU (02/C05-i01.02/2022.PC644943391-00000051); and in part by the Fundação para a Ciência e a Tecnologia (FCT), I.P., under Project UIDB/04524/2020; and in part by the Scientific Employment Stimulus-Institutional Call under Grant CEECINST/00051/2018.

\bibliographystyle{IEEEtran}
\bibliography{Bibliography/bibliography.bib}

\end{document}